\def\year{2016}\relax
\documentclass[letterpaper]{article}
\usepackage{aaai16}
\usepackage{times}
\usepackage{helvet}
\usepackage{courier}
\usepackage{url}
\usepackage{graphicx}
\usepackage{graphicx}
\usepackage{epstopdf}

\begin{document}
\title{A Markov Chain based Ensemble Method for Crowdsourced Clustering{}}
\author{Sujoy Chatterjee$^1$, Enakshi Kundu$^1$ and Anirban Mukhopadhyay$^1$\\
$^1$Department of Computer Science and Engineering, University of Kalyani, Nadia -- 741235, India\\
E-mail: \{sujoy, anirban\}@klyuniv.ac.in, enakshikundu@ymail.com\\
}

\maketitle

\begin{abstract}
In presence of multiple clustering solutions for the same dataset, a clustering ensemble approach aims to yield a single clustering of the dataset by achieving a consensus among the input clustering solutions. The goal of this consensus is to improve the quality of clustering. It has been seen that there are some image clustering tasks that cannot be easily solved by computer. But if these images can be outsourced to the general people (crowd workers) to group them based on some similar features, and opinions are collected from them, then this task can be managed in an efficient manner and time effective way. In this work, the power of crowd has been used to annotate the images so that multiple clustering solutions can be obtained from them and thereafter a Markov chain based ensemble method is introduced to make a consensus of multiple clustering solutions.
\end{abstract}

\section{Introduction}
Clustering is a common unsupervised learning method, which is used to find hidden patterns or groupings in data. These groups are termed as clusters, and there are different types of clustering techniques \cite{Pavel2002,jain:99} that partition the dataset in different ways. In unsupervised classification, known as clustering, it is not known beforehand how the data is grouped. 
There are some drawbacks in all existing clustering techniques. Few clustering methods are also very sensitive to the initial clustering settings. 
In cluster analysis, the evaluation of results is generally done using of cluster validity indexes \cite{mukhopadhyay2015,daviesd1979,rand1971,dunn1974,hubert:1985} which are employed to measure the quality of clustering results. However, there is no cluster validity index that impartially evaluates the results of any clustering algorithm. So, to combine the multiple diverse clustering solutions for achieving an improved clustering, an ensemble of clustering solutions is needed.

Although over the years numerous clustering ensemble algorithms \cite{strehl2002,ayad2008,chatterjee2013} have been proposed to solve different issues related to cluster analysis. These algorithms have several benefits and pitfalls. Moreover, there are some hard image clustering tasks that cannot be solved by a computer in limited amount of time. But if this large image clustering task can be outsourced to numerous crowd workers \cite{Brabham:2013,Hovy:2013,Lease:11} and clustering solutions can be obtained from them, then the task can be solved in a very effective way  through a cluster ensemble approach. 

In this paper, an online platform is designed to collect the clustering solutions from the crowd workers over some tricky images and a Markov chain based ensemble technique is proposed to find a robust consensus from multiple crowd based clustering solutions. This proposed scheme might be a better utilization of enormous human power that can easily solve the large image clustering task.

\section{Problem Formulation}
Let $Z=\{z_1, z_2, \ldots, z_o\}$ be a set of $o$ data objects and there are $p$ crowd workers each of whom provides an individual clustering solution. So, $E=\{e_1, e_2, \ldots, e_p\}$ denotes a set of $p$ input clustering solutions obtained from them. Now in this problem, the number of clusters is assumed to be fixed in all the clustering solutions. So, the objects are partitioned into $n$ clusters denoted as $C= \{c_1, c_2, \ldots, c_n\}$. The objective of the problem is to find out the ensemble solution $\tau$ from these multiple input clustering solutions $E=\{e_1, e_2, \ldots, e_p\}$ so that the similarity between $\tau$ and all of the input clustering solutions of $E$ is maximized.

\section{Proposed Model}
\begin{figure}[!h]
\centering
\includegraphics[width =7cm, height=4.3cm ]{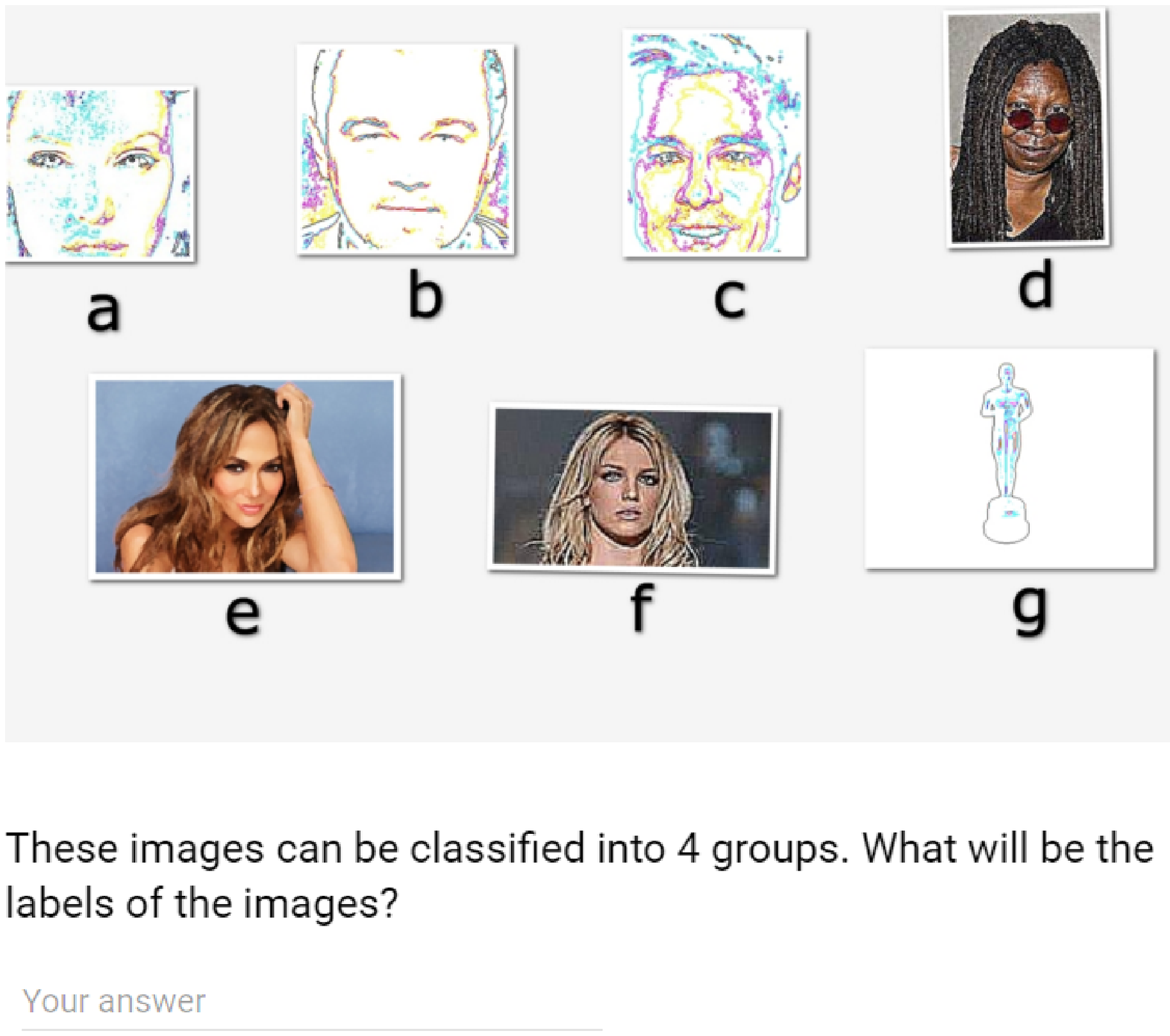}
\caption{Snapshot of the first question posted to crowd workers. }
\label{fig:3}
\end{figure}

We have made an online platform where we have posted some tricky images (that is hard for a computer to distinguish) and solicited crowd opinions to cluster those images based on their similarity. As this is a fixed size clustering problem so the number of clusters has also been posted. As various crowd workers might group the images from different viewpoints, diverse clustering solutions can be obtained. In this way, the artificial dataset is created. Fig. \ref{fig:3} shows the snapshot of a question for image clustering task posted to the crowd workers. Here, the question contains 7 images (a, b, c, d, e, f, g) which are asked to be clustered into 4 groups based on some similarities in features. After obtaining the clustering solutions from them, the proposed ensemble method is applied in order to make more robust consensus clustering solution.

\subsection{Markov Chain based Ensemble Technique}
Cluster labels given by the crowd workers are very symbolic, i.e., two identical clustering solutions might appear different for using different cluster labels. To resolve this label correspondence problem, here, we have chosen the standard label to be that of a reference clustering solution (a clustering solution which is most similar to the rest of the other solutions) by using Adjusted Rand Index (ARI) \cite{hubert:1985} as a similarity measure. The solution assigned the maximum weight (based on ARI), is chosen as the reference partition $e_r$ according to whose labeling all the other clustering solutions are relabeled.

The clustering ensemble problem can be solved using a Markov chain that is specified by a set of states $N=\{1,2, \ldots, n\}$ and a $n \times n$ transition matrix $T$, each of whose entries is a non negative real number in [0, 1] representing a probability. To solve this problem, object-wise transition matrix is formed and  the following steps are carried out to find the consensus solution.

\begin{itemize}
\item {\bf Step 1.} In the first step, the number of possible states is found. Here, the possible states mean the unique clusters. So, if there are $n$ clusters in the clustering solutions, the number of states will be $n$. 

\item {\bf Step 2.} The weight of each clustering solution is measured by the average similarity of it with rest of the solutions in terms of ARI. %Now, let $w = \{ w_{e_1}, w_{e_2}, \ldots , w_{e_p}\}$ be the weights of the $p$ clustering solutions. 
Here normalized weight $w^\prime = \{ w_{e_1}^\prime, w_{e_2}^\prime, \ldots , w_{e_p}^\prime\}$ is used for $p$ clustering solutions.
% where each $w_{e_i}^\prime =  \frac{w_{e_i}}{\sum\limits_{i=1}^p w_{e_i}}, \forall i $. 

\item {\bf Step 3.} We construct a transition matrix $ T$ of order $n \times n$, whose elements $T_{ij}$ denotes the probability of transition from state $i$ to state $j$ for each value $i, j$ of data object $y \in Z$. At first the transition matrix is initialized with zeros but it is then modified as described below.% In this context, the cell $i, j$ means the labeling $L(e_r)$, $L(e_s)$ given by two clustering solutions $e_r$ and $e_s$, respectively.

%Let there are $p$ clustering solutions. So the probability of getting the label provided by any solution from any of the $p$ solutions is $1/p$. 
To compute the different cell values of the transition matrix (for a particular object) from a set of input clustering solutions, the label given by one clustering solution is taken initially as reference and the label provided by other solutions are considered. In this step, the normalized weight of the other clustering solutions are summed up to compute the transition probability.

Now this step is computed $\forall e_m \in E$ keeping the reference clustering solution $e_k$ fixed. In this way, after completion of this step for a particular reference clustering solution, another clustering solution is taken as the reference and the rest of the solutions are considered for it to determine the values of the transition matrix. Note that, now as all the labels are already standardized, to form the transition matrix, label of each of the clustering solution should be compared with label of rest of the solutions.

\item {\bf Step 4.} Make the transition matrix $T$ ergodic as follows: 
\begin{center}
$T_{ij}=  T_{ij} + \frac{1-\Sigma_{j=1}^n  T_{ij}}{n}$, $\forall i,j \in \{1, 2,  \ldots, n\}$ 
\end{center}

\item {\bf Step 5.}  Define a stationary distribution array $sd_y$ of order $1$ $\times$ $n$, where $n$ denotes the number of clusters. Initially, the distribution is taken as uniform distribution. Then  $sd_y*T$ is computed for $m$ times so that $sd_y$ reaches a converging point. Finally the cluster label for which the distribution becomes maximum is selected as the final cluster label of the corresponding object.
\end{itemize}
In this way, the same steps are repeated for all the objects and final clustering solution is achieved.

\section{Experimental Design and Results}
We performed experiments on real-life datasets (obtained from UCI Machine Learning Repository) to find the efficacy of the proposed method and compared it with that of four well-known existing cluster ensemble algorithms, namely CSPA, HGPA, MCLA \cite{strehl2002} and BCE \cite{Wang:2010}. For image clustering task 25 crowd workers provided their solutions. The adopted performance metrics are ARI, Rand Index (RI) \cite{rand1971}, Hubert Index (HI) and Mirkin Index (MI) \cite{hubert:1985}. It is evident from Table \ref{tab:5} that the proposed algorithm provides good performance and thus it produces better consensus from multiple crowdsourced clustering solutions. 

\begin{table}[!h]
\small
\renewcommand{\arraystretch}{1.0}%
\centering
\caption {Performance metric values for balance scale dataset.} \label{tab:5}% The best scores over a column are shown in bold.} 
 \begin{tabular}{|c | c | c | c | c | c | c | c | c |} 
 \hline
 Algorithm & Adjusted Rand & Rand & MI & HI  \\ [0.5ex] 
 \hline
  CSPA &	0.1579 &	0.5941 &	0.4059 &	0.1883 \\
 \hline
 HGPA &	0.1482 &	0.5216 &	0.4784 &	0.0431   \\
 \hline
 MCLA  &	0.0118 &	0.5985 &	0.4015 &	0.1971  \\  
 \hline
BCE    & 0.0830	&0.5616	&0.4384	&0.1232			\\ 
 \hline
Proposed &	\textbf{0.1767} &	\textbf{0.6071} &	\textbf{0.3929} &	\textbf{0.2141} \\ 
 \hline
\end{tabular}
\end{table}

\section{Conclusions}
In this paper, a crowdsourcing model for grouping a set of tricky images is introduced and a Markov chain based ensemble method has been proposed. It can be adopted as an effective tool to achive a good consensus from multiple diverse clustering solutions. Furthermore, it can also be extended to work with input crowdsourced clustering having variable  number of clusters instead of fixed number of clusters considering other features (e.g., confidence and bias) of crowd workers.

\section*{Acknowledgements}
The authors would like to thank the anonymous reviewers for their valuable comments that greatly helped to improve the quality of the paper. All the authors would like to thank the crowd contributors involved in this work.

\bibliographystyle{aaai}
\bibliography{References}

\end{document}